\documentclass[aps,prd,twocolumn,reprint,showpacs,superscriptaddress,nofootinbib,amsfonts,amssymb,amsmath]{revtex4-1}  
\usepackage[english]{babel}
\usepackage[utf8]{inputenc}
\usepackage{graphicx}  
\usepackage{dcolumn}   
\usepackage{bm}        
\usepackage{amssymb}   
\usepackage{setspace}
\usepackage[hidelinks]{hyperref}
\usepackage{verbatim}
\usepackage{indentfirst}
\usepackage{textcomp}
\usepackage{braket}
\usepackage{amsmath}
\usepackage{blindtext}
\usepackage{xcolor}
\usepackage{amsfonts}
\begin{document}
\author{Valerio Cascioli}
\affiliation{Department of Physics, University of Rome “La Sapienza”, P.le Aldo Moro 5, Roma, 00185, Italy}
\affiliation{INFN Roma1, Rome, Italy}
\author{Giovanni Montani}
\affiliation{Department of Physics, University of Rome “La Sapienza”, P.le Aldo Moro 5, Roma, 00185, Italy}
\affiliation{ENEA C. R. Frascati, Via E. Fermi 45, Frascati, Roma 00044, Italy}
\author{Riccardo Moriconi}
\affiliation{Department of Physics, University of Napoli "Federico II", via Cinthia, Napoli, I-80216, Italy}
\begin{abstract}
We develop a suitable technical algorithm to implement a separation of the Minisuperspace configurational variables into quasi-classical and purely quantum degrees of freedom, in the framework of a Polymer quantum Mechanics reformulation of the canonical dynamics. We then implement the obtained general scheme to the specific case of a Taub Universe, in the presence of a free massless scalar field. In particular, we identify the quasi-classical variables in the Universe volume and a suitable 
function of the scalar field, while the purely quantum degree of freedom corresponds to the Universe anisotropy. We demonstrate that the Taub cosmology is associated to a cyclical Universe, oscillating between a minimum and maximum volume turning points, respectively. The pure quantum Universe anisotropy has always a finite value. 
\end{abstract}
\title{WKB approximation for the Polymer quantization of the Taub Model}
\maketitle

\section{Introduction}
One of the most puzzling shortcomings of the Universe representation in modern Cosmology is the presence of an initial singularity, predicted by the Einstein equation, but undoubtedly it is an unphysical ultraviolet divergence to be somehow regularized \cite{montani2, kolb, weinberg}. 

Various non-singular cosmological models can be constructed on a classical and quantum level, see for instance \cite{cianfrani1} but the emergence of a Bounce Cosmology can be attributed to the implementation of Loop Quantum Gravity on a cosmological setting, see \cite{ashtekar1}. When a metric approach is considered, the most natural way to deal with a singularity-free cosmological model, relies on the implementation of a Polymer Quantum Mechanics approach to the Minisuperspace \cite{corichi, ashtekar1}. This approach is, de facto, a discretization procedure of the considered configurational variables (in cosmology they are Universe scale factors), which turn out to live on a graph and can have only a discrete spectrum, for a picture of the literature in merit, see \cite{montani1, montaniP, montaniP3, corichi}.

From the side of the quantum physics of space-time, an highly non-trivial question concerns the absence of a parametric (external) time variable, when the canonical method is implemented \cite{isham, rovelli, montaniNP, montaniD}. 

Among many different proposal to construct a suitable clock in quantum gravity \cite{kuchar}, it stands the WKB approach proposed in \cite{vilenkin}, see also \cite{kiefer}. The proposed scenario relies on a Born-Oppenheimer approximation, in which some Minisuperspace variables behaves slowly and are quasi-classical degrees of freedom, becoming a good clock for the fully quantum and rapidly changing variables. In other words, the time dependence of the wave function of the quantum part is recovered by its dependence on the quasi-classical variables, in turn linked to the coordinate time.

The present work explores the possibility to deal with a cosmological model in which the singularity is regularized via a Polymer Quantum Mechanics approach and a time dependence of the Universe wave function is defined via a Born-Oppenheimer decomposition of the quantum dynamics. The non trivial technical question we address here is to reconcile the momentum representation of the quantum dynamics, mandatory for a Polymer quantization, as developed in \cite{corichi} for the continuum limit and the WKB scheme, thought in the coordinate representation. The crucial point is that the potential term emerging in the Minisuperspace model is, in general, non quadratic in the configurational variables, like instead in general is the Kinetic part of the Hamiltonian in the momenta. To overcome this difficulty, we introduce a suitable and general algorithm and then we implement it in the particular and important case of a Taub Cosmological model \cite{taub, montani2}. 

The classical Taub solution links a non-singular expanded universe to a singular point of the space-time curvature, as it naturally arises because it is nothing more than a Bianchi IX model with two equal cosmic scale factors (the spatial geometry is the same of a closed Robertson-Walker geometry).

The cosmological model resulting from our regularization is a very intriguing paradigm: we get an evolutionary quantum picture, whose description corresponds to a (non-singular) cyclical Universe. 

Our study of the Taub cosmology in the presence of a scalar field is performed using Misner-Chitrè-like variables \cite{imponente}. The quasi-classical variables are identified in the scalar field and in the one that is most directly linked to the Universe volume, actually in the adopted variables the isotropic metric component and the anisotropies are somehow mixed together. The quantum degree of freedom is identified in the relic anisotropy coordinate of the Taub model, a suitable redefinition of the variables is also necessary during the technical derivation. 

The resulting evolutionary (Schrodinger) equation for this anisotropy variable has, in the spirit of the analysis here presented, two main physical implications: i) the Taub model is reduced to a cyclical Universe, evolving between a minimum and a maximum value of the Universe variables, offering an intriguing paradigm for the physical implementation of a cosmological history: clearly the maximum volume turning point is expected to live in a classical domain of the Universe dynamics, while the Bounce turning point has a pure quantum character , in the sense of a Polymer regularization; ii) the Universe anisotropy is always finite in value as a result of the singularity regularization and its specific value in the Bounce turning point depends on the initial conditions of the system, but in principle, it can be restricted to small enough values to make the Bounce dynamics unaffected by their behavior, i.e. the applicability of the Born-Oppenheimer approximation is ensured in the spirit of the analysis provided in \cite{agostini}. 

The paper is structured as follows: two section in order to introduce the Minisuperspace, the Bianchi Models and the Polymer Quantum Mechanics, one section to generalize the Vilenkin approach in both the representations, a section in which we will implement the generalized approach to the Taub model in both the Classical and Polymer Quantum Mechanics, at the end there will be a section where we discuss the obtained result and the conclusions.

\section{Minisuperspace and Bianchi Models}
The idea of the Minisuperspace was born from the possibility of reducing the general problem of the quantum gravity to the simple case of a space-time highly symmetric, with a dynamics in a finite dimensions scheme, and the quantization to a natural Dirac's prescription for the Universe wave function. 

For the purpose of this paper we will limit ourselves to the Homogeneous Universes that are described by the Bianchi Models; those models represent all the possible universes that are homogeneous but anisotropic. There are 9 different models but the most studied are Bianchi I, V and IX, that contain respectively the Flat, the Open and the Close FRW model once taken the isotropic limit.

Alexander Vilenkin chose to study the dynamics of the Primordial Universe in a Minisuperspace scheme, so that the the metric depends only on the coordinates and the wave function is approximated by its WKB expansion where the classical variables are treated differently from the quantum ones. The Ukraine physicist studied a method relatively elegant and linear that allows us to obtain a probabilistic interpretation of the wavefunction of the Universe, in accordance with the studies of Bohr.

Vilenkin studied the simple case of a homogeneous Minisuperspace, in which the variables are the tri-geometries, and he chose to use the ADM formulation. From the Super Hamiltonian constraint derives the Wheeler-De Witt (WD) equation, that is simply a generalized n-dimensional Klein-Gordon (KG) equation with variable mass. From this equation it derives a conserved current that assures the preservation of the probability. In particular the WD equation is: 
\begin{equation}
\label{eq:1}
\left[ g^{\alpha \beta } \left( \frac{ \partial S}{ \partial p_{ \gamma} } \right) p_ {\alpha } p_{\beta} - U \left( \frac{ \partial S}{ \partial p_{ \gamma} } \right) - H_q \right] \psi \left( p \right) = 0\ . 
\end{equation}
The Action \emph{S} is in the ADM form and the wave function will be: 
\begin{equation}
\psi \left(p\right) = A \left(p \right) e^{\frac{i}{\hslash} S \left(p \right)} \phi \left(p,q \right) \ .
\end{equation}

We will apply this method (once generalized) to a particular Bianchi IX Universe: The Taub Universe. This model allows us to restrict the problem to that of a particle in one dimension that hits a potential wall with only a degree of freedom, corresponding to take a preferential direction in the bi-dimensional phase space of Bianchi IX, in particular we choose to eliminate one of the anisotropic variables of the model. \cite{montani2}

\section{Polymer Quantum Mechanics}
The Polymer Quantum Mechanics is an alternative representation of the Quantum Mechanics, it's based on the Weyl Commutation Rules (WCR) that differ greatly from the classical ones (CCR). The WCR state that for the quantization of the system the commutator between two canonical variables becomes $\left[\hat{x},\hat{p} \right] = i\hslash \cos \left(\mu p \right)$ and so the algebra is quite difficult. 

One of the feature of this representation is that at the beginning of the study one has to decide which variables are discrete and which are not, so fundamentally one has to introduce a lattice structure in the system and see where it leads. There are two possibilities of implementation for these requirements, depending on the choice of the polarization of the wave functions, namely the \emph{q-polarization} and the \emph{p-polarization}; for the purpose of this paper we choose the latter because the equations are easier to study. 

The problem is that when one associates a discrete character to one of the variables, the Weyl algebra assures that the operator associated to its conjugate variable doesn't exist. This creates a lot of  problems when one tries to quantize the system, one of the most serious is that it's necessary to decide a range of reliability in which is possible to approximate that operator and never leave it. 

We can start by defining abstract kets $\ket{\mu}$ labeled by a real number. These shall belong to the Hilbert space $\mathcal{H} _{\operatorname{poly}}$. From these states, we define a generic state that correspond to a choice of a finite collection of numbers $\mu _i \in \mathbb{R} $ with $i=1,2,\dots ,N$. Associated to this choice, there are $N$ vectors $\ket{\mu _i}$, so we can take a linear combination of them $\ket{\psi} = \sum ^{N} _{i=1} a_i \ket{\mu _i}$. The fundamental kets are orthonormal and the Polymer Hilbert Space $\mathcal{H} _{\operatorname{poly}}$ is non-separable.

There are two basic operators on this Hilbert Space, namely the \emph{label} operator $\hat{\epsilon}$ and the \emph{displacement} operator $\hat{s} \left(\lambda \right)$ \cite{corichi} and based on the chosen polarization they can be assigned to the classical operators. In our polarization we have to approximate the momenta in the Hamiltonian and in order to do that we can define 
\begin{equation}
p \approx \frac{1}{\mu} \sin \left( \mu p \right) \ , \ \ \ p^2 \approx \frac{2}{\mu ^2} \left[1- \cos \left( \mu p \right) \right] \ , 
\end{equation} 
valid in the regime $p\ll 1/\mu $.

One can ask why this representation is so important, the answer is that there are systems which don't admit a standard description and if we can define a continuum limit of the Polymer then it's possible to quantize the system with this representation and find its dynamics, at least approximately. Studying systems with exact solutions like the harmonic oscillator and the free particle, it has been established that the results given by both representation are exactly the same once the continuum limit of the Polymer Quantum Mechanics is taken.

There are, indeed, great expectations for this new representation, especially in Quantum Cosmology where all the new theories (such as the Loop Quantum Cosmology) suggest the existence of a finite inferior limit of the volume of the Universe and so the presence of a lattice structure in the space-time.

\section{Generalized Vilenkin Approach}
\label{GVA}
In this paragraph I will extend the study of Vilenkin \cite{vilenkin} to the case of a totally general homogeneous universe. We will start from the Wheeler-De Witt equation in the momenta base that is written as: 
\begin{equation}
\left[ g^{\alpha \beta } \left( \frac{ \partial S}{ \partial p_{ \gamma} } \right) p_ {\alpha } p_{\beta} - U \left( \frac{ \partial S}{ \partial p_{ \gamma} } \right) - H_q \right] \psi \left( p \right) = 0
\end{equation}
and the Action \emph{S} is in the ADM form and the wave function will be: 
\begin{equation}
\psi \left(p\right) = A \left(p \right) e^{\frac{i}{\hslash} S \left(p \right)} \phi \left(p,q \right) \ .
\end{equation}

The first step to achieve the generalized approach is to introduce a generalization of the Derivative operator that will greatly help in the following. Let's start from the simplest cases: 
\begin{equation}
\begin{split}
D^{\mu} _{p} \left[p^{\nu} \right] &= \frac{\Gamma \left( \nu +1 \right)}{\Gamma \left( \nu - \mu +1 \right)} p^{\nu - \mu} \\
D^{\mu} _{p} \left[e^p \right] &= D^{\mu} _{p} \left[ \sum ^{\infty} _{k=0} \frac{p^k}{k!} \right] = \sum ^{\infty} _{k=0} \frac{p^{k- \mu}}{\Gamma \left(k+1- \mu\right)} \equiv E^{p} _{\mu}\ ;
\end{split}
\end{equation}
where $E^{p} _{\mu} $ is the generalized exponential function defined by:
\begin{equation}
\label{expgen}
\begin{split}
E^{ap} _{\mu} & \equiv p^{\mu} e^{ap} \gamma ^{*} \left(\mu, ap\right)\ ,\\
\gamma ^{*} \left(\mu, ap\right) & \equiv e^{-ap} \sum ^{\infty} _{j=0} \frac{\left(ap \right) ^j}{\Gamma \left(\mu +j+1 \right)}\ .
\end{split}
\end{equation}
Let's further advance and introduce the case of a grade-\emph{n} polynomial as the exponent
\begin{equation}
D^{\mu} _{p} \left[e^{ap} \right] \!=\! D^{\mu} _{p} \! \left[ \sum ^{\infty} _{k=0} \frac{\left(ap \right)^k}{k!} \right]\! \!=\! a^{\mu}\! \sum ^{\infty} _{k=0} \frac{\left(ap\right) ^{k- \mu}}{\Gamma \left( k+1 -\mu \right)}\! \equiv \! a^{\mu} E^{ap} _{\mu}
\end{equation}  
Let's define another function for the purpose of this paper
\begin{equation}
Ln_{\mu} \left[ E^{ap} _{\mu} \right] \equiv ap\ .
\end{equation}

When everything is taken into account it must be said that as soon as we put a generic function in the place of the exponential of a polynomial, all the maths starts to decade because the initial definition has a lot of problems that are solved only in the case of polynomial functions. In the following this generalized derivative will be often used because we will only consider functions that are related to polynomial.

\subsection{Ordinary Case}
\label{CFG}
The Hamilton-Jacobi equation is described by the first order expansion of \eqref{eq:1}. In order to obtain it, it's necessary to expand the exponential in its power series and take only the right order terms. We get: 
\begin{equation}
\psi \left(p\right) = A \left(p \right) \left[1 +\frac{i}{\hslash}S \left(p\right) - \frac{1}{2 \hslash ^2} S^2 \right] \phi \left(p,q \right)
\end{equation}
and so equation \eqref{eq:1} becomes at the lowest order: 
\begin{equation}
\begin{split}
 g^{ \alpha \beta } \left( \frac{ \partial S}{ \partial p_{ \gamma} } \right) p_ {\alpha} p_{\beta} A \left(p \right) \left( - \frac{1}{2 \hslash ^2} S^2 \right) \phi \left(p,q \right) + \\
- U \left( \frac{ \partial S}{ \partial p_{ \gamma} } \right) A\left(p \right) \phi \left(p,q \right) = 0\ ;
\end{split}
\end{equation}
with the due simplifications and introducing the notation $\left( \frac{ \partial}{ \partial p_{ \gamma} } \right) \equiv \left( \partial _{\gamma} \right)$ we obtain
\begin{equation}
\label{HJO}
g^{\alpha \beta } \left( \frac{ \partial S}{ \partial p_{ \gamma} } \right) p_{\alpha} p_{\beta} \frac{S^2}{\hslash ^2 } + 2U \left( \frac{ \partial S}{ \partial p_{ \gamma} } \right) = 0\ .
\end{equation}
that reproduce exactly the Hamilton-Jacobi equation of the classical case once we identify $\left( \frac{ \partial S}{ \partial p_{ \gamma} } \right)$ with $h^{\gamma} $. 

At the next order we get two separate equations given that, as in the case analyzed by Vilenkin, we can exploit the adiabatic approximation. Let's start analyzing first the equation for the amplitude $A$ and then the one for the quantum wavefunction.
Studying the general case, we don't have the explicit forms of the metric and the potential term, and so we can't let them act directly on the wavefunction; what we can do is, instead, multiply by the identity both of the terms defining 
\begin{equation}
\mathbb{I} = \left(i\hslash \partial _{\gamma} \right) ^{-1} \left(i \hslash \partial _{\gamma}\right) \ .
\end{equation}

The desired equation can be obtained at the next order of the expansion in $\hslash$. Multiplying by the identity defined above and having the exotic derivative acting only on the amplitude while the normal one acts on the exponential term we obtain: 
\begin{equation}
g^{\alpha \beta} p_{\alpha} p_{\beta} \left[\! \left( \partial ^{-1} A\right) \! \left( \partial \  e^{\frac{i}{\hslash} S} \right)\! \right] \phi - U \left[\! \left(\partial A \right)\! \left(\partial ^{-1} e^{\frac{i}{\hslash} S} \right)\! \right] \phi = 0
\end{equation} 
Those are not the only terms at the right order so we multiply again the equation by the identity and we get:
\begin{equation}
\label{eqA}
\begin{split}
i g^{\alpha \beta} p_{\alpha} p_{\beta} \left\lbrace \!\left[2 \partial ^{-1} \!\left( \partial \partial ^{-1} \!A \right)\! \left( \partial S \right)\! \right] + \left[\! \left( \partial ^{-2}\! A \right)\! \left( \partial ^2\! S \right)\! \right] \right\rbrace e^{\frac{i}{\hslash} S}\! + \\
-U\hslash \left[2 \left(\partial A \right) \left(\partial ^{-1} e^{\frac{i}{\hslash} S} \right) \right] = 0
\end{split}
\end{equation}
And this is the equation for the amplitude $A\left(p \right)$ .

Now we analyze the equation for the pure quantum wavefunction.
As in the above case, we multiply the initial equation by the identity, but this time the important part, in order to obtain the  Schr\"odinger equation, is when the exotic derivative acts on the exponential and the normal one acts on the quantum term; before we approach the real calculation it's opportune expanding the action in its power series: $S\left(p \right) = \sum^{\infty} _{k=0} c_k \left(t\right) p^k \left(t\right) $.

Let's start applying the exotic derivative on the exponential using the definition: 
\begin{equation} 
\label{dergene}
\begin{split}
\left(i \hslash \partial _{\gamma} \right) ^{-1} e^{\frac{i}{\hslash} \sum^{\infty} _{k=0} c_k p^k} = \left( i\hslash \partial _{\gamma} \right) ^{-1} \prod ^{\infty} _{k=0} \sum ^{\infty} _{j=0} \left[\frac{\left( \frac{i}{\hslash} c_k p^k \right) ^j}{j!} \right]= \\
=\frac{1}{i\hslash}  \left\lbrace \frac{1}{\sum ^{\infty} _{k=1} \left[\frac{i}{\hslash} c_k k \right] } \prod ^{\infty} _{k=0} \sum ^{\infty} _{j=0} \frac{\left(\frac{i}{\hslash}c_k p^k \right)^{j+1}}{\Gamma \left(j+2 \right)}\right\rbrace = \\
= \frac{1}{i\hslash \mathfrak{C}} \prod ^{\infty} _{k=0} E_{-1} ^{\frac{i}{\hslash} c_k p^k} \ ,  
\end{split}
\end{equation}
where $\Gamma$ is the Euler Gamma Function, for the sake of notation I have defined $\mathfrak{C} = \sum ^{\infty} _{k=1} \frac{i}{ \hslash} c_k$ and I have introduced the generalized exponential function defined above in \eqref{expgen}. The equation 
\begin{equation}
\label{eqschr}
\hat{H} _0 \left( \partial ^{-1} e^{ \frac{i}{\hslash} S} \right) \left( \partial \phi \right) = \hat{H} _q e^{\frac{i}{\hslash} S} \phi
\end{equation}
becomes, exploiting \eqref{dergene}, 
\begin{equation}
\label{eq5}
\hat{H_0} \frac{1}{ \mathfrak{C}} \prod ^{\infty} _{k=0} E_{-1} ^{\frac{i}{\hslash} c_k p^k} \left( \partial _{\gamma} \phi \right) = \hat{H_q} e^{\frac{i}{\hslash} \sum ^{\infty} _{k=0} c_k p^k} \phi\ ,
\end{equation}
where $\hat{H_0}$ is the classical part of the WD. We highlight in particular the property of one of the terms in equation \eqref{eq5}: 
\begin{equation}
\frac{\frac{1}{ \mathfrak{C}} \prod ^{\infty} _{k=0} E_{-1} ^{\frac{i}{\hslash} c_k p^k}}{e^{\frac{i}{\hslash} \sum ^{\infty} _{k=0} c_k p^k}} = \frac{F\left(p\right)}{F'\left(p\right)} = \frac{1}{\partial _{\gamma} Ln_{-1} \left[F\left(p\right)\right]}
\end{equation}
and so we obtain:
\begin{equation}
\hat{H_0} \frac{1}{\partial _{\gamma} Ln_{-1} \left[F\left(p\right)\right]} \left( \partial _{\gamma} \phi \right) = \hat{H_q} \phi\ .
\end{equation}
We can rewrite the p-derivative of the logarithm as its time derivative times $\frac{\partial t}{\partial p_{\gamma}} $, thanks to the properties of the differentials, and it ensures that it's possible to obtain the time derivative even of the quantum terms. Let's see how it can be done
\begin{equation}
\hat{H_0} \frac{1}{\partial _{t} Ln_{-1} \left[F\left(p\right)\right] \partial _{\gamma} t} \left( \partial _{\gamma} \phi \right) = \hat{H_q} \phi\ .
\end{equation}

The time derivative of the logarithm, $\partial _{t} Ln_{-1} \left[F\left(p\right)\right]$ can be written as $\frac{i}{\hslash} \mathfrak{D}$ and so we obtain the equation:
\begin{equation}
\frac{1}{\frac{i \mathfrak{D}}{\hslash}} \left( \frac{\partial p_{\gamma}}{\partial t} \frac{\partial \phi}{\partial p_{\gamma}} \right) = \hat{H_0} ^{-1} \hat{H_q} \phi \ .
\end{equation}

We can take all the temporal dependence of the above equation and define a new time derivative in $\tau$ in order to get 
\begin{equation}
-i\hslash \frac{1}{\mathfrak{D}} \left( \frac{\partial \phi}{\partial t} \right) =  \hat{H_0} ^{-1} \hat{H_q} \phi \ ,
\end{equation}
if we define $\tau $ such that 
\begin{equation}
\frac{\partial}{\partial \tau} \equiv \frac{1}{\mathfrak{D}} \left(\frac{\partial }{\partial t} \right)
\end{equation}
and after we take all the other terms to the second member we obtain:
\begin{equation}
\label{eqshr}
-i \hslash \frac{\partial \phi}{\partial \tau} = \hat{H_1} \phi
\end{equation}
that is the desired Schr\"odinger's equation for the quantum wavefunction. In equation \eqref{eqshr} I have defined $$\hat{H} _1 \equiv \hat{H} ^{-1} _0 \hat{H} _q $$.

\subsection{Polymer Case}
As seen in the section about the Polymer quantum mechanics, imposing a Polymer quantization means assuming a discrete structure for some of the variables of the phase space. The consequence of this fact is that it's not possible to associate to the conjugated variables quantum differentials operators as in the ordinary case. The Polymer paradigm, to solve this problem, consists in the substitution  $p \rightarrow \frac{1}{\mu} \sin \left(\mu p \right)$. As a consequence, the Polymer version of the WD equation is:   
\begin{equation}
\left[ \frac{\hslash ^2}{\mu ^2} \hat{g^{\alpha \beta}} \sin \left( \frac{\mu p_{\alpha}}{\hslash} \right) \sin \left( \frac{\mu p_{\beta}}{\hslash} \right) - U -H_q \right] \psi = 0\ .
\end{equation}
Expanding it at the lowest order and using the power series of the exponential we find the Hamilton-Jacobi equation for the Polymer case: 
\begin{equation}
\begin{split}
- \frac{\hslash ^2}{\mu ^2} \hat{g^{\alpha \beta}} \sin \left( \frac{\mu p_{\alpha}}{\hslash} \right) \sin \left( \frac{\mu p_{\beta}}{\hslash} \right) \frac{AS^2 \phi}{2 \hslash ^2} - UA\phi &=0 \\
\Rightarrow \frac{1}{\mu ^2} \hat{g^{\alpha \beta}} \sin \left( \frac{\mu p_{\alpha}}{\hslash} \right) \sin \left( \frac{\mu p_{\beta}}{\hslash} \right) S^2 + 2U =0 \ .
\end{split}
\end{equation}

As seen above at the next order we find two separate equations because of the adiabatic approximation. In order to find those equations we will use the same method of the last section with the identity defined by $$ \mathbb{I} = \left(i\hslash \partial _{\gamma Pol} \right) ^{-1} \left(i \hslash \partial _{\gamma Pol}\right) \ .$$ Using the same notation of the last section we get: 
\begin{widetext}
\begin{equation}
\begin{split}
&\frac{\hslash ^2}{\mu ^2} \hat{g^{\alpha \beta}} \sin \left( \frac{\mu p_{\alpha}}{\hslash} \right) \sin \left( \frac{\mu p_{\beta}}{\hslash} \right) \phi -U  \mathbb{I} \psi=0  \ \ \ \Rightarrow \\
&\frac{\hslash ^2}{\mu ^2} \hat{g^{\alpha \beta}} \sin \left( \frac{\mu p_{\alpha}}{\hslash} \right) \sin \left( \frac{\mu p_{\beta}}{\hslash} \right)  \mathbb{I}  \left( \partial ^{-1} _{Pol} A \right) \left(\partial _{Pol} E_+ \right) \phi - U  \mathbb{I} \left[ \left( \partial _{Pol} A \right) \left(\partial ^{-1} _{Pol} E_+ \right) \right] \phi =0 \\
&\Rightarrow \frac{\hslash ^2}{\mu ^2} \hat{g^{\alpha \beta}} \sin \left( \frac{\mu p_{\alpha}}{\hslash} \right) \sin \left( \frac{\mu p_{\beta}}{\hslash} \right) \left( \partial ^{-1} _{Pol} \right) \left(\partial _{Pol} \right) \left[ \left( \partial ^{-1} _{Pol} A \right) \left(\partial _{Pol} E_+ \right) \right] \phi -  U  \left( \partial ^{-1} _{Pol} \right) \left(\partial _{Pol} \right)  \left[ \left( \partial _{Pol} A \right) \left( \partial ^{-1} _{Pol} E_+ \right) \right] \phi =0 \ .
\end{split}
\end{equation}
If we write explicitly the known terms we obtain: 
\begin{equation}
\begin{split}
 \frac{i\hslash }{\mu ^2} \hat{g^{\alpha \beta}} \sin \left( \frac{\mu p_{\alpha}}{\hslash} \right) \sin \left( \frac{\mu p_{\beta}}{\hslash} \right) \left[ 2 \left( \partial ^{-1} _{Pol} A \right) \left( \partial _{Pol} S \right) + \left( \partial ^{-2} _{Pol} A\right) \left( \partial ^2 _{Pol} S \right) \right] E_+ - U \hslash \left[2 \left( \partial _{Pol} A \right) \left(\partial ^{-1} _{Pol} E_+ \right) \right] = 0 
\end{split}
\end{equation}
This is the equation for the Polymer amplitude $A$. 
\end{widetext}

Although the calculation made till now demonstrates that the equations that we obtain in both the representations are the same taking into account the correction introduced by the passage from one to the other, let's see what happen to the quantum wavefunction. 
The method is exactly the same of the last section since $f\left[ \sin \left(p\right) \right] \equiv f\left(p\right)$. 
The equation
\begin{equation}
\hat{H} _{0_{Pol}} \left( \partial ^{-1} _{Pol} e^{ \frac{i}{\hslash} S} \right) \left( \partial _{Pol} \phi \right) = \hat{H} _{q_{Pol}} e^{\frac{i}{\hslash} S} \phi
\end{equation}
becomes, exploiting \eqref{dergene}, 
\begin{equation}
\label{eq25}
\hat{H} _{0_{Pol}} \frac{1}{ \mathfrak{C}} \prod ^{\infty} _{k=0} E_{-1} ^{\frac{i}{\hslash} c_k p^k} \left( \partial _{\gamma} \phi \right) = \hat{H} _{q_{Pol}} e^{\frac{i}{\hslash} \sum ^{\infty} _{k=0} c_k p^k} \phi\ ,
\end{equation}
where with $\hat{H} _{0_{Pol}}$ has been indicated the classical part of the WD in the Polymer representation. In this particular case the generalized exponential function contains all the Polymer correction and it is substantially different from the ordinary one. We highlight, even in this case, the property of one of the terms in equation \eqref{eq25}: 
\begin{equation}
\frac{\frac{1}{ \mathfrak{C}} \prod ^{\infty} _{k=0} E_{-1} ^{\frac{i}{\hslash} c_k p^k}}{e^{\frac{i}{\hslash} \sum ^{\infty} _{k=0} c_k p^k}} = \frac{F\left(p\right)}{F'\left(p\right)} = \frac{1}{\partial _{\gamma} Ln_{-1} \left[F\left(p\right)\right]}
\end{equation}
and so we get:
\begin{equation}
\hat{H} _{0_{Pol}} \frac{1}{\partial _{\gamma } Ln_{-1} \left[F\left(p\right)\right]} \left( \partial _{\gamma} \phi \right) = \hat{H} _{q_{Pol}} \phi\ .
\end{equation}
Taking into account the properties of the differentials, we can rewrite the p-derivative of the logarithm as the time derivative of the logarithm times $\frac{\partial  t}{\partial p_{\gamma}} $ \begin{equation}
\hat{H} _{0_{Pol}} \frac{1}{\partial _{t} Ln_{-1} \left[F\left(p\right)\right] \partial _{\gamma} t} \left( \partial _{\gamma} \phi \right) = \hat{H} _{q_{Pol}} \phi\ .
\end{equation}

The time derivative of the logarithm, $\partial _{t} Ln_{-1} \left[F\left(p\right)\right]$ can be written as $\frac{i}{\hslash} \mathfrak{D} _{Pol}$ and so we get the equation:
\begin{equation}
\frac{1}{\frac{i \mathfrak{D} _{Pol}}{\hslash}} \left( \frac{\partial p_{\gamma}}{\partial t} \frac{\partial \phi}{\partial p_{\gamma}} \right) = \hat{H} ^{-1} _{0_{Pol}} \hat{H} _{q_{Pol}} \phi \ .
\end{equation}

We can take all the temporal dependence of the above equation and define a new time derivative in $\tau$ in order to get  
\begin{equation}
-i\hslash \frac{1}{\mathfrak{D} _{Pol}} \left( \frac{\partial \phi}{\partial t} \right) =  \hat{H} _{0_{Pol}} ^{-1} \hat{H} _{q_{Pol}} \phi \ ,
\end{equation}
if we define $\tau _{Pol} $ such that 
\begin{equation}
\frac{\partial}{\partial \tau _{Pol}} \equiv \frac{1}{\mathfrak{D} _{Pol}} \left(\frac{\partial }{\partial t} \right)
\end{equation}
and after we take all the other terms to the second member we obtain:
\begin{equation}
\label{eqshrpol}
-i \hslash \frac{\partial \phi}{\partial \tau _{Pol}} = \hat{H} _{1_{Pol}} \phi \ ,
\end{equation}
where $$\hat{H} _{1_{Pol}} \equiv \hat{H} ^{-1} _{0_{Pol}} \hat{H} _{q_{Pol}} $$.

The equation above is the desired Schr\"odinger equation and it's equivalent to the ordinary case. Clearly both in the time variable and in the terms of the Hamiltonian there is the Polymer correction, but formally they are the same.

\subsection{Conserved Current}
We analyze now the probability current defined from the equation \eqref{eq:1} in order to obtain the continuity equation that allow us to replicate the Vilenkin approach. We start from  $\psi \left(p\right) = A\left(p\right) e^{\frac{i}{\hslash}S\left(p\right)} \phi \left(p,q\right)$ and its complex conjugated  $\psi ^{*} \left(p\right) = A^{ \dagger} \left(p\right) e^{-\frac{i}{\hslash}S\left(p\right)} \phi ^{*} \left(p,q\right)$. 
Imposing the Hamiltonian constraint we can formally find $p_{\alpha} = f\left(h^{\alpha} \right)$ along the equation of motion. Furthermore it is possible to use the Hamilton equations to find the analytical expressions for $\dot{p}$ and $\dot{h} $.
The definition of the probability current is: 
\begin{equation}
\label{defcorr}
J^{\delta} = \frac{i}{2} \hslash  p_{\alpha} p_{\beta} \hat{g}^{\alpha \beta} \left( \frac{\partial}{\partial p_{\delta}} \right) ^{-1} \left( \frac{\partial}{\partial p_{\gamma}} \right)^{-1}\left[ \psi ^{*} \partial _{\gamma} \psi - \partial _{\gamma} \psi ^{*} \psi \right]
\end{equation}

We differentiate the above equation to obtain:
\begin{widetext}
\begin{equation}
\begin{split}
\partial _{\delta} J ^{\delta} = \frac{i}{2} &\hslash  p_{\alpha} p_{\beta} \hat{g}^{\alpha \beta} \left[ \left( \partial _{\gamma} ^{-1} \psi ^{*} \right) \left( \partial _{\gamma}  \psi \right) -\left( \partial _{\gamma} \psi ^{*} \right) \left( \partial _{\gamma} ^{-1} \psi  \right)- \psi ^{*} \psi + \psi ^{*} \psi  - \psi ^{*} \psi + \psi ^{*} \psi -  \psi ^{*} \psi + \psi ^{*} \psi\right]+ \\
+ \frac{i}{2} \hslash  \partial _{\delta} \left(p_{\alpha} p_{\beta} \hat{g}^{\alpha \beta} \right)& \partial _{\delta} ^{-1} \left[ \left( \partial _{\gamma} ^{-1} \psi ^{*} \right) \left( \partial _{\gamma}  \psi \right) -\left( \partial _{\gamma} \psi ^{*} \right) \left( \partial _{\gamma} ^{-1} \psi  \right)\right]+ \frac{i}{2} \hslash  p_{\alpha} p_{\beta} \hat{g}^{\alpha \beta} \partial _{\delta} ^{-1} \left\lbrace \partial _{\delta} \left[ \left( \partial _{\gamma} ^{-1} \psi ^{*} \right) \left( \partial _{\gamma}  \psi \right) -\left( \partial _{\gamma} \psi ^{*} \right) \left( \partial _{\gamma} ^{-1} \psi  \right)\right]\right\rbrace + \\
&+ \frac{i}{2} \hslash  p_{\alpha} p_{\beta} \hat{g}^{\alpha \beta} \partial _{\delta} ^{-1} \partial _{\gamma} ^{-1} \left[ \left( \partial _{\delta} \psi ^{*} \right) \left( \partial _{\gamma} \psi \right) - \left( \partial _{\gamma} \psi ^{*} \right) \left( \partial _{\delta} \psi \right) +  \psi ^{*} \left( \partial _{\delta} \partial _{\gamma} \psi \right) - \left( \partial _{\delta} \partial _{\gamma} \psi ^{*} \right) \psi \right]\ ; \\
\end{split}
\end{equation}
with the due simplifications and defining
\begin{equation}
\Lambda \equiv \left[ \left( \partial _{\gamma} ^{-1} \psi ^{*} \right) \left( \partial _{\gamma}  \psi \right) -\left( \partial _{\gamma} \psi ^{*} \right) \left( \partial _{\gamma} ^{-1} \psi  \right) \right] 
\end{equation} 
we obtain the following equation:
\begin{equation}
\label{eqmostro}
\begin{split}
\partial _{\delta} J^{\delta}\! &=  \frac{i}{2} \hslash  p_{\alpha} p_{\beta} \hat{g}^{\alpha \beta}\! \left[4\Lambda \!+\! \left( \partial _{\delta} ^{-1} \partial _{\gamma} ^{-1} \psi ^{*} \right)\! \left( \partial _{\delta} \partial _{\gamma} \psi \right) \!  -\! \left( \partial _{\delta} \partial _{\gamma}  \psi ^{*} \right)\! \left(\partial _{\delta} ^{-1} \partial _{\gamma} ^{-1} \psi \right)\!+\! \left( \partial _{\delta} \partial _{\gamma} ^{-1} \psi ^{*} \right)\! \left( \partial _{\delta} ^{-1} \partial _{\gamma} \psi \right)\! - \!\left( \partial _{\delta} ^{-1} \partial _{\gamma}  \psi ^{*} \right)\! \left(\partial _{\delta}  \partial _{\gamma} ^{-1} \psi \right)\! \right]+ \\
& + \frac{i}{2} \hslash ^2 \partial _{\delta}\! \left( p_{\alpha} p_{\beta} \hat{g}^{\alpha \beta} \right)\! \partial _{\delta} ^{-1} \Lambda \ .
\end{split}
\end{equation}
\end{widetext}
The last two terms within the square brackets of the above equation are  null for the properties of the generalized derivative while the last line of the right hand side reproduce exactly the equation of motion and so it's null. 

From the analysis of the term in $\Lambda $  it is evident that the only terms at the right order in $\left(\hslash \right)$ are: 
\begin{equation}
\begin{split}
&\Lambda = i \left( \partial _{\gamma} ^{-1} \vert A \vert ^2 \right) \left( \partial _{\gamma} S \right) \vert \phi \vert ^2 + \\
&+\!\vert A \vert ^2\! \left( \partial _{\gamma} ^{-1} E_{-} \right)\! E_{+} \phi ^{*} \!\left( \partial _{\gamma} \phi \right) - \vert A \vert ^2 E_{-} \!\left( \partial _{\gamma} ^{-1} E_{+} \right)\! \left( \partial _{\gamma} \phi ^{*} \right)\!  \phi\ ,
\end{split}
\end{equation}
with the notation $E_{\pm} \equiv e^{ \pm \frac{i}{\hslash} S}$.
A property very important of the generalized derivative is, as in the ordinary one, the Leibniz law, that applied in this case gives the relation
\begin{equation}
\left( \partial _{\gamma} ^{-1} E_{-} \right)\! E_{+} + E_{-}\! \left( \partial _{\gamma} ^{-1} E_{+} \right)\! =\! D_{p} ^{-1}\! \left( E_{-} E_{+} \right)\! =\! D_{p} ^{-1} \left(1 \right) = p
\end{equation}
and so it is possible to express one term of the left hand side as a function of the other, in order to maintain the initial ordering we choose the relation $E_{-} \left( \partial _{\gamma} ^{-1} E_{+} \right) = p - \left( \partial _{\gamma} ^{-1} E_{-} \right) E_{+} $ and we get
\begin{equation}
\Lambda = i \left( \partial _{\gamma} ^{-1} \vert A \vert ^2 \right) \left( \partial _{\gamma} S \right) \vert \phi \vert ^2 + \vert A \vert ^2 \left( \partial _{\gamma} ^{-1} E_{-} \right) E_{+} \left( \partial _{\gamma} \vert \phi \vert ^2 \right)
\end{equation}
the term that contains $p$ is of a different order and so it can be neglected.

As for the second term on the right hand side of the first line of the equation \eqref{eqmostro} the only term of the right order is $ i \left( \partial _{\delta} ^{-1} \partial _{\gamma} ^{-1} \vert A \vert ^2 \right) \left( \partial _{\delta} \partial _{\gamma} S \right) \vert \phi \vert ^2$.
At the end we can say that the dominant terms of the equation \eqref{eqmostro} reduce to:
\begin{equation}
\begin{split}
\partial _{\delta} J^{\delta} \!&=\! i\! \left[\! \left( \partial _{\gamma} ^{-1} \vert A \vert ^2 \right)\! \left( \partial _{\gamma} S \right)\! \vert \phi \vert ^2\! +\! \left( \partial _{\delta} ^{-1} \partial _{\gamma} ^{-1} \vert A \vert ^2 \right)\! \left( \partial _{\delta} \partial _{\gamma} S \right)\! \vert \phi \vert ^2 \right]\! +\\ 
& + \vert A \vert ^2 \left( \partial _{\gamma} ^{-1} E_{-} \right) E_{+} \left( \partial _{\gamma} \vert \phi \vert ^2 \right)\ .
\end{split}
\end{equation}
Those are the equations \eqref{eqA} and \eqref{eqschr} for the Universe wavefunction and for its complex conjugate derived before. Considering their definitions the term on the right hand side it's identically null and so even in the case of this study there is a conserved probability current. This demonstration is valid for both Standard and Polymer Quantum mechanics once taken the correct assumptions. 

\section{Application to the Taub Model}
In this section I will applicate the results of the previous sections to the Taub Model (one of the particular cases of Bianchi IX model), the result will be a quantum wavefunction for the Universe that will allow us to infer the behavior of the Early Universe. 

Although usually the best choice for this kind of study are the Misner Variables $\left( \alpha, \beta_{+}, \beta_{-} \right)$ for their immediate physical interpretation: $\alpha$ is related to the volume of the Universe, while the $\beta$ are related to the two physical degree of freedom of the Gravitational Field, for the following discussion I chose another set of variables more complicate and with a not immediate physical sense, the Misner-Chitrè variables. They enable us to study the dynamics of the system in the so-called \emph{Poincaré Half Plane} that eliminate the dynamics of the potential's wall. In particular the two set of variables have the following relations \cite{moriconi}: 
\begin{equation}
\begin{split}
\alpha - \alpha_0 &= -e^{\tau} \frac{1+u+u^2+v^2}{\sqrt{3}v}\\
\beta_{+} &= e^{\tau} \frac{-1+2u+2u^2+2v^2}{2\sqrt{3}v}\\
\phi &= e^{\tau} \frac{-1-2u}{2v}\ .
\end{split}
\end{equation}
In order to make the Vilenkin Approach works it's necessary to insert a term of matter, for the purpose of this study I chose the Scalar Field. 

The dynamics of this model near the singularity reduces to the one of a particle that hit continuously the walls of a pseudo-triangular box \cite{misner4} \cite{misner5}; the cosmological singularity is reached when the trajectory ends in one of the corner of the box. This model consists in taking one preferential direction in the $\beta $-plane, and so only one of the walls of the Bianchi IX Universe that the particle hits only one time and then goes directly in the opposite corner. This means that the Misner $\beta _- $ is identically null and so the Misner-Chitrè $u$ is always a constant and equal to $-1/2$, implying that the conjugate momentum $p_u$ is always zero.

In the chosen variables the Super-Hamiltonian constraint $\mathcal{H} = 0$ leads to a WD equation without all the terms in $p_u$. In this case the metric assumes the simple form 
\begin{equation}
\label{metric1}
ds^2 = \frac{\epsilon}{v^2} \left[du^2 + dv^2 \right]\ .
\end{equation}
In order to make the math easier we change again variables, introducing 
\begin{equation}
\begin{split}
v &= \rho \sin \left(2\delta \right) \\
u &= \rho \cos \left(2 \delta \right) \ ,
\end{split}
\end{equation}
with $0< \rho < \infty $ and $0< \delta < \pi $. If we insert them in the metric it's simple to verify that \eqref{metric1} becomes 
\begin{equation}
\label{metric2}
ds^2 = \epsilon \left[\frac{d\rho ^2}{\rho^2 \sin ^2 \left(2\delta \right)} + \frac{8 d \delta ^2}{\sin ^2 \left(2\delta \right)} \right]\ .
\end{equation}

If now we define $ dx = d\rho / \rho$ e $ d \theta = d\delta / \sin \left(2\delta \right)$ and integrate them we find two variables with the same limits of the Misner-Chitrè ones 
\begin{equation}
\begin{split}
x &= \log \vert \rho \vert \ \ , \ \ -\infty < x < \infty \\
\theta &= \frac{1}{2} \log \vert \tan \left( \delta \right) \vert \ \ , \ \  -\infty < \theta < \infty \ ;
\end{split}
\end{equation}
with a few calculations it's possible to rewrite the term $\sin ^2 \left( 2 \delta \right)$ present in \eqref{metric2} as a function of the new variable $\theta$ only as 
\begin{equation}
\begin{split}
\boxed{\sin ^ 2 \left( 2 \delta \right)} =\  &4 \sin ^2 \left( \delta \right) \cos ^2 \left( \delta \right) = \\
4 \sin ^2 \left[ \arctan \left( e^{2\theta} \right) \right] & \cos ^2 \left[ \arctan \left( e^{2\theta} \right) \right]= \\
4 \frac{e^{4\theta}}{e^{4\theta}+1} \frac{1}{e^{4\theta}+1} &= \boxed{\frac{1}{ \cosh ^2 \left(2\theta \right)}}
\end{split}
\end{equation}
where I used the relation $\sin^2 \left[\arctan \left(x \right) \right] = \frac{x^2}{x^2+1}$ and the definition of the hyperbolic cosine. With these substitutions the metric becomes 
\begin{equation}
ds^2 = \epsilon \left[ \frac{ d\rho ^2}{ \cosh ^2 \left( 2\theta \right)} + 8 d \theta ^2 \right] \ ,
\end{equation}
we can choose a gauge and we decided to use the condition $H' = \theta H$ and so the Hamiltonian of the system becomes
\begin{equation}
\label{Hamiltoniana}
H = \theta \left[ -p_{\tau} ^2 - \frac{p_{\theta} ^ 2}{8} + \cosh ^2 \left(2\theta\right) p_x ^2 \right]\ .
\end{equation}

\subsection{Ordinary Case}
Let's analyze this Hamiltonian \eqref{Hamiltoniana} in order to get the  equations for the dynamics of the system, we derive them via the Ehrenfest Theorem as
\begin{equation}
\label{EHO1}
\begin{split}
\langle \dot{p_\theta} \rangle &= \frac{1}{i\hbar} \langle \left[ p_\theta , H  \right] \rangle = p^2 _\tau + \frac{p^2 _\theta}{8} \\
\langle \dot{ \tau} \rangle &= \frac{1}{i\hbar} \langle \left[ \tau , H  \right] \rangle = -2\theta p_{\tau} \\
\langle \dot{ \theta} \rangle &= \frac{1}{i\hbar} \langle \left[ \theta , H  \right] \rangle = -\frac{\theta p_{\theta} }{4} \\
\langle \dot{ x} \rangle &= \frac{1}{i\hbar} \langle \left[x , H  \right] \rangle = 2\theta \cosh ^2 \left(2\theta  \right) p_{x} 
\end{split}
\end{equation}

The Hamiltonian \eqref{Hamiltoniana} doesn't depend explicitly on $\tau$ and $x$ and so their momenta are constants of motion.
Those are the equations that describe the dynamics of the Universe.

Now I will adapt the Vilenkin approach to the Taub Universe. First of all I will use a wavefunction in the form $\psi \! \left( p_{\tau}, p_{\theta}, p_x \right) = A\! \left( p_{\tau}, p_{\theta} \right) e^{\frac{i}{\hbar}S} \chi \! \left( p_{\tau}, p_{\theta}, p_x \right) $ where $S \! \left( p_{\tau}, p_{\theta} \right) $ is the Action of the system.  If we take the lowest order of the Hamiltonian constraint $H\psi = 0$ we find the Hamilton-Jacobi equation for the system as in \eqref{HJO}: 
\begin{equation}
\label{HJo}
p_{\tau} ^2 S \operatorname{d} \! S - \frac{\hbar ^2 p_{\theta} \operatorname{d} \! p_\theta}{4} = 0\ ;
\end{equation}
as seen in the previous section we derive the equations for the Amplitude of the wavefunction and the Schrodinger equation for the dynamics of the quantum  variables respectively as: 
\begin{equation}
\label{Ao}
-\frac{i}{\hbar} p_{\tau} ^2 A \frac{\partial S}{\partial p_\theta} + \frac{p^2 _\theta}{8}\frac{\partial A}{\partial p_\theta}  = 0
\end{equation}
\begin{equation}
\label{So}
i\hbar \frac{\partial \chi}{\partial t} = p_x ^2 \chi 
\end{equation}
Putting together equations \eqref{HJo} and \eqref{Ao} we get the amplitude of the Universe wavefuntion as $A= A_0 e^{-4i \frac{p_\tau}{p_\theta}}$ and this completely characterize the classical part of the probability density defined above.

The variable $t$, that appears in \eqref{So}, is a time-variable defined by $ \frac{\partial }{\partial t} \equiv  \frac{1}{\frac{\partial S}{\partial p_\theta} \cosh^2 \left(2\theta \right)} \frac{\partial}{\partial z} $ and $z$ is the Vilenkin time defined by $\frac{\operatorname{d}}{\operatorname{d} \! z} \equiv \dot{p_\tau} \frac{\partial}{\partial p_\tau} + \dot{p_\theta} \frac{\partial}{\partial p_\theta}$. If we consider a quantum part of the Universe wavefunction in the form $\chi = e^{\frac{i}{\hslash} Et} \phi \! \left(p_{\theta}, p_x \right)$ and we put it in \eqref{So} we can solve it and we find
\begin{equation}
\begin{split}
&E = \frac{p_x ^2}{2} \\
&\phi \! \left( p_x \right) = C_1 \delta \! \left(p_x - p_{E,x} \right) + C_2 \delta \! \left(p_x + p_{E,x} \right) \\
&\phi \! \left( x \right) = \frac{1}{\sqrt{2\pi}} e^{-ip_{E,x} x} \left(C_1 + C_2 e^{2ip_{E,x} x}\right) \ .
\end{split}
\end{equation} 

\subsection{Polymer Case}
Let's go back to the Hamiltonian \eqref{Hamiltoniana} and use the Polymer Quantum Dynamics instead of the classical one. If we want the Hamilton equations we must remember that in this case the canonical commutator is $\left[\hat{x_i},\hat{p_i} \right] = i\hslash \cos \left(\mu p_i \right)$. The Wheeler-De Witt equation in this case is in the form:
\begin{equation}
\label{WDWP}
\begin{split}
\theta \left\lbrace -\frac{1}{\mu ^2} \sin ^2 \left( \mu p_{\tau} \right)  -\frac{1}{8 \mu ^2} \sin ^2 \left( \mu p_{\theta} \right)\right\rbrace \Psi + \\
+\theta \left\lbrace  \frac{\cosh ^2 \left(2\theta \right)}{\mu ^2} \sin ^2 \left(\mu p_x\right) \right\rbrace \Psi = 0 
\end{split}
\end{equation}
With the same calculations of the previous section we find the equations for the dynamics of the particle Universe 
\begin{equation}
\label{EHP}
\begin{split}
\langle \dot{p_\theta} \rangle &= \frac{1}{\mu ^2} \sin ^2 \left( \mu p_{\tau} \right) + \frac{1}{8 \mu ^2} \sin ^2 \left( \mu p_{\theta} \right) \\
\langle \dot{ \tau} \rangle &= \frac{\theta}{\mu} \sin \left(2\mu p_{\tau} \right) \ \\\
\langle \dot{ \theta} \rangle &= \frac{\theta}{4\mu} \sin \left(2\mu p_{\theta} \right) \ \ \\
\langle \dot{x} \rangle &= \frac{2\theta \cosh ^2 \left(2\theta \right)}{\mu} \sin \left(2\mu p_{x} \right)\ \ . 
\end{split}
\end{equation}
As in the previous section, even in this case the other two momenta are constants of motion.
Those are the equations that describe the dynamics of the Early Universe. 

Now we use the Vilenkin approach in this case, from equation \eqref{WDWP} we can derive the Hamilton-Jacobi equation, the equation for the amplitude of the wavefunction and the Schrodinger equation for the quantum variables respectively 
\begin{equation}
\label{HJop}
\frac{1}{\mu ^2} \sin ^2 \left(\mu p_{\tau} \right) S \operatorname{d} \! S -\frac{\hslash ^2}{4\mu ^2} \sin  \left(\mu p_{\theta} \right) \cos  \left(\mu p_{\theta} \right) \operatorname{d} \! p_\theta = 0 
\end{equation}
\begin{equation}
\label{Aop}
\begin{split}
\frac{i A \sin ^2 \left(\mu p_\tau \right) }{\hbar}  \frac{\partial S}{\partial p_\theta} + \frac{\sin ^2 \left(\mu p_\theta\right)}{8}\frac{\partial A}{\partial p_\theta} =0
\end{split}
\end{equation}
\begin{equation}
\label{SoP}
i\hbar \frac{\partial \chi}{\partial t_{pol}} = \frac{\left[1-\cos \left(\mu p_x \right) \right] \chi }{\mu ^2}
\end{equation}
Using together equations \eqref{HJop} and \eqref{Aop} we get the amplitude of the Universe wavefuntion as $A= A_0 e^{-4i \frac{\sin \left(\mu p_\tau \right)}{\sin \left( \mu p_\theta \right)}}$ and this completely characterize the classical part of the probability density defined above.

The variable $t_{pol}$, that appears in \eqref{SoP}, is a time-variable defined by $ \frac{\partial }{\partial t_{pol}} \equiv  \frac{\sin \left(\mu p_\tau \right)}{\hbar \cos \left( \mu p_\theta \right) \cosh^2 \left(2\theta \right)} \frac{\partial }{\partial z_{pol}} $ where $z_{pol}$ is the Vilenkin time in the Polymer representation. If we consider a quantum part of the Universe wavefunction in the form $\chi = e^{\frac{i}{\hslash} kt_{pol}} \phi \! \left(p_{\theta}, p_x \right)$ and we put it in \eqref{SoP} we can solve it and we find
\begin{equation}
\label{SHRO}
\begin{split}
&k= k\! \left(\mu \right) = \frac{1}{\mu ^2} \left[1-\cos \left(\mu p_x \right) \right] \leq k_{max} =  \frac{2}{\mu ^2} \\
& \phi _{k, \mu} \! \left(p_x \right) = C_1 \delta \! \left(p_x - p_{k, \mu} \right) + C_2 \delta \! \left(p_x + p_{k, \mu} \right) \\
&\phi _{k, \mu} \! \left(x \right) = \frac{1}{\sqrt{2\pi}} e^{-ip_{k, \mu} x} \left(C_1 + C_2 e^{2ip_{k, \mu} x} \right) \ .
\end{split}
\end{equation}
We can notice that those are the same results of the previous section once taken into account the Polymer modifications, moreover we can also notice that the eigenvalue here has an upper limit and this will be very important in the dynamics of the Universe. 

\section{Discussion}
\begin{figure}[h]
\includegraphics[width = \columnwidth]{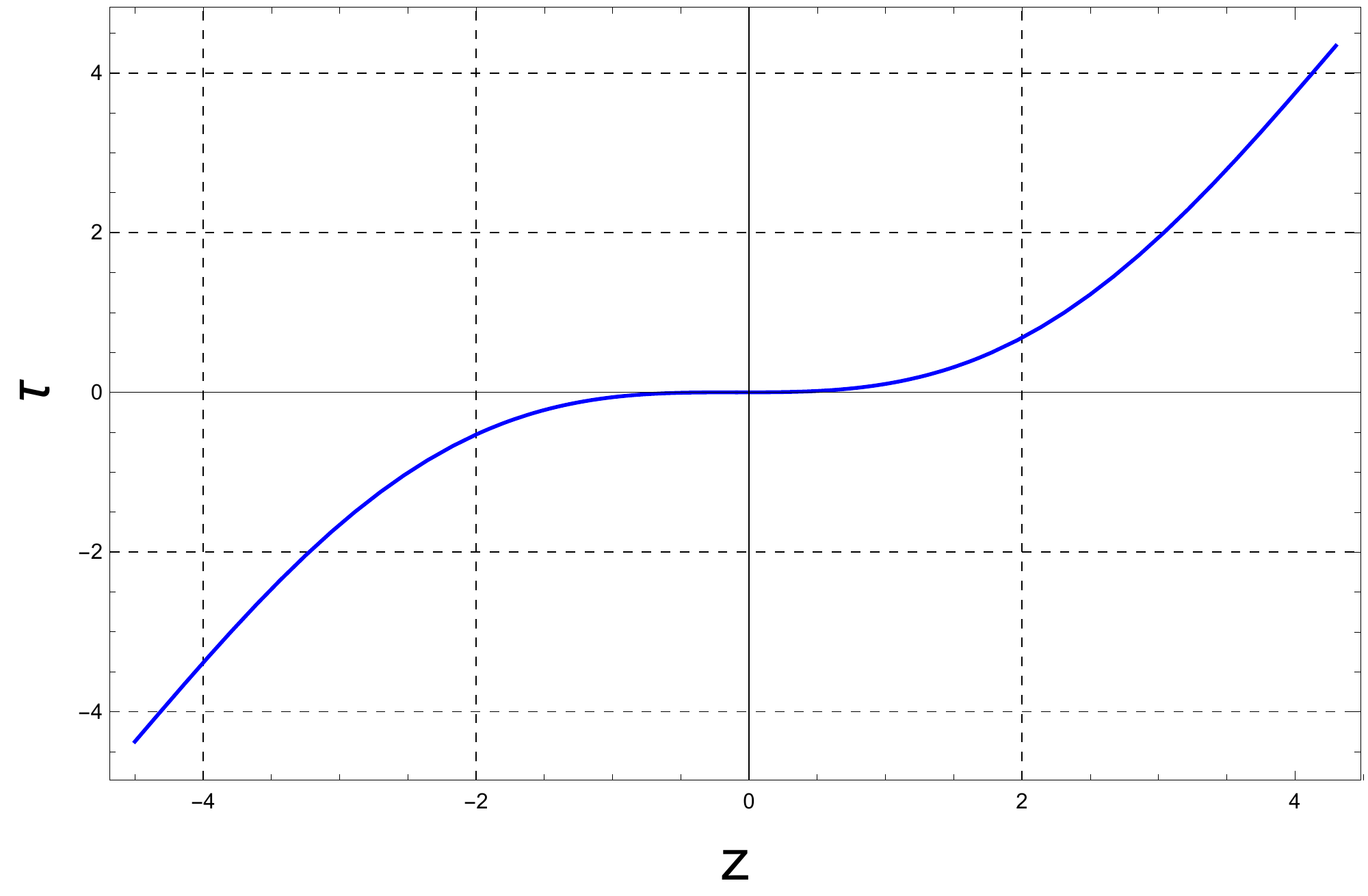}
\caption{Dynamics of the volume of the Universe in the Vilenkin time variable z defined above. The Universe starts at finite volume, reach the potential barrier (z=0) and then goes toward the singularity of the model (z = $\infty$).} \label{figclassv}
\end{figure}

\begin{figure}[h]
\includegraphics[width = \columnwidth]{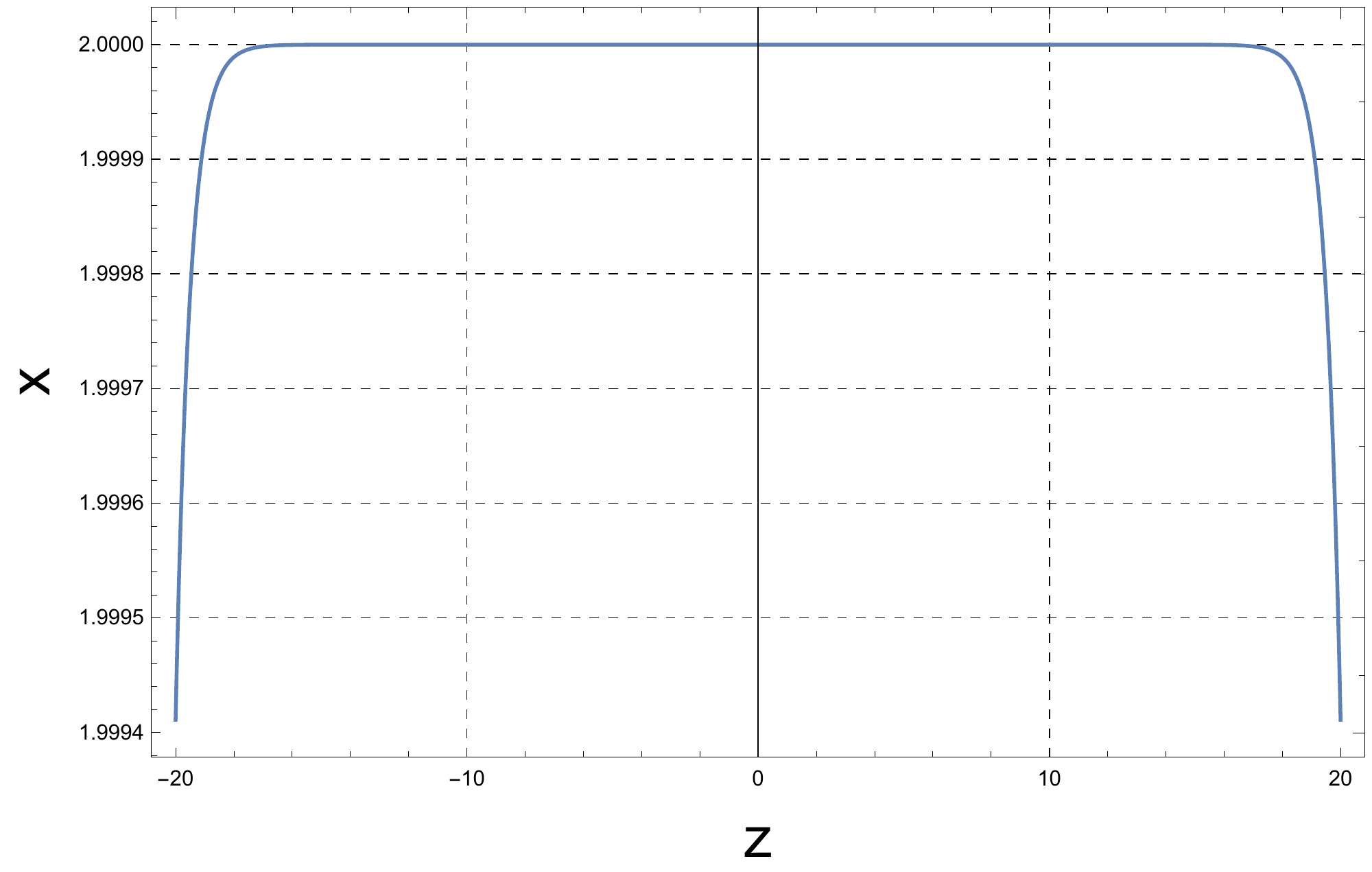}
\caption{Dynamics of the anisotropies of the universe in the Vilenkin time variable z defined above. The Universe starts with a finite degree of anisotropy, it then reaches a constant value near the potential wall (z=0) and then explodes in the singularity of the model (z = $\infty$).} \label{figclassa}
\end{figure}

We now analyze the equations that we found in the previous section, in particular the Hamilton equations \eqref{EHO1} and \eqref{EHP} obtained in the two different cases. If we integrate those systems we obtain the following equations for the volume of the Universe $\tau $, the scalar field $\theta $  and its momenta $p_\theta$
\begin{equation}
\label{HC}
\begin{split}
\langle p_\theta \rangle &= 2\sqrt{2} p_\tau \tan \left(J \right) \\
\langle \theta \rangle &= C_2 \cos ^{2.}  \left(J \right) \\
\langle  \tau \rangle &= C_3 + \frac{4\sqrt{2}C_2}{3} \cos ^{3.} \left(J \right) \csc \left(J \right) \cdot \\
& \cdot \ _2  F _1 \left[\frac{1}{2}, \frac{3}{2}, \frac{5}{2}, \cos ^{2.} \left(J \right)  \right] \sin \left(J \right)\ . \\
\end{split}
\end{equation} 
Those are the equation for the ordinary case in which we defined $J= \frac{1}{4} \left(p_\tau z + 8\sqrt{2} C_1\right)$, the dynamics of the volume of the Universe will be plotted in Fig \eqref{figclassv} while the equation for the anisotropy $x$ is numerically solved and we will show its dynamics in Fig \eqref{figclassa}. The Universe starts at a point with finite volume, evolves towards the potential wall and then goes straight into the singularity without the possibility to evade it. The anisotropies, instead, explode near the singularity and are practically null near the wall. 

\begin{figure}[h]
\includegraphics[width = \columnwidth]{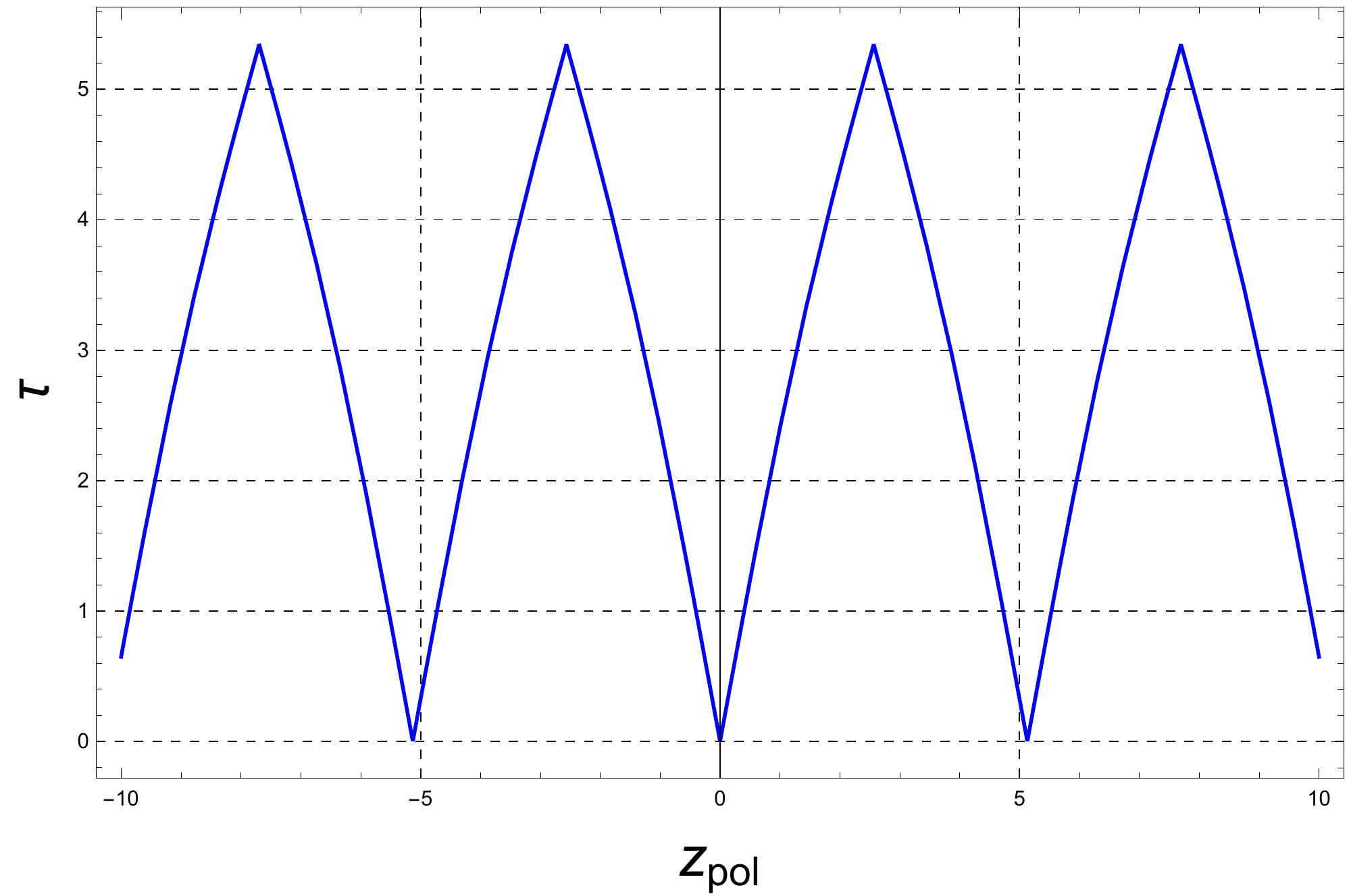}
\caption{Dynamics of the volume of the universe in the Vilenkin time variable $z_{pol}$ defined above. The Universe starts with a finite volume, it then reaches a series of turning points.} \label{figpolv}
\end{figure}

\begin{figure}[h]
\includegraphics[width = \columnwidth]{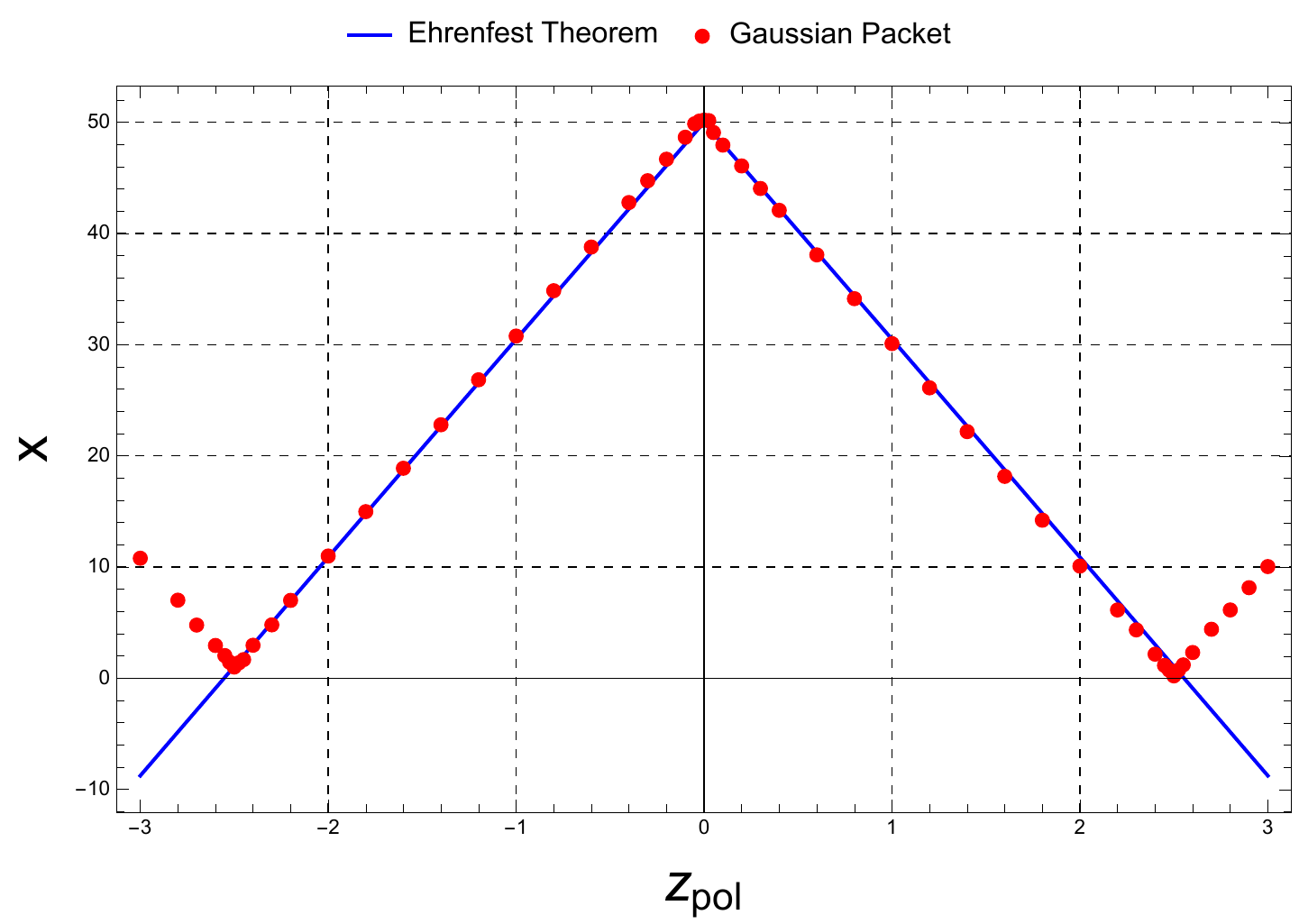}
\caption{Dynamics of a Gaussian packet build from the solution of the Schrodinger equation in the Vilenkin time variable compared to the dynamics obtained with the Ehrenfest Theorem. } \label{figpola}
\end{figure}

For the Polymer case there are no analytical solutions of the system \eqref{EHP}, all the equations are numerically integrated and their dynamics will be plotted in Fig \eqref{figpolv} and Fig \eqref{figpola} (solid line). 

 The plots shown in Fig \eqref{figpolv} and Fig \eqref{figpola} allow us to state that the Taub model can be reduced to a singularity-free model with a cyclical behavior in both volume and anisotropies. In the four plots it's possible to highlight the main differences between the two representations, in the ordinary case the singularity is unavoidable, while in the Polymer approach there is a periodic behavior of the Universe variables, and so the singularity is regularized. 

If we take the general solution of the Schrodinger equation \eqref{SHRO} with the boundary conditions due to the Taub Cosmological Model, that in our variables it can be shown that read as $\phi \! \left( x_0 \right) \equiv \phi \! \left( \infty \right) \equiv 0$ (where we defined $x_0 = \log \left(\frac{1}{\sqrt{2}} \right)$), the wave function of the Universe becomes 
\begin{equation}
\Psi = \frac{C_1}{\sqrt{2\pi}} e^{ \frac{i}{\hslash}kt_{pol} } \left[e^{ipx}-e^{ip\left(x_0 -x \right)} \right] 
\end{equation}
With this we can now construct a Gaussian packet and study its dynamics, the first step is to define the packet as 
\begin{equation}
\Xi = \int ^{k_{\operatorname{Max}}} _{0} \exp \left[-\frac{\left(k-k_0 \right)^2}{2\sigma ^2}\right] \Psi \operatorname{d} \! k
\end{equation}
then we numerically evaluate this integral at different times in order to obtain a dynamics of the Gaussian packet, in figure \eqref{figpola} we have shown the results of our analysis and we have even compared the evolution of the Gaussian packet with the dynamics of the mean value of the quantum anisotropy that we got with the Ehrenfest theorem. 

From the plot it can be seen that the two trajectories coincide up to the Bounce-turning-points, then the packet dynamics reveals a series of turning point like the volume variable and we can even see a correspondence between the behavior of the two variables. In those points the variance, calculated via the distribution theory, on the Ehrenfest equation for the anisotropy is comparable to the mean value and so we can say that in the bounce-turning-points of this model, the Ehrenfest theorem cannot be very descriptive and so the packet dynamics shows the correct evolution of the anisotropy variable. Thus we can conclude that in our approach the true singularity of the Taub Model is regularized with the Polymer Quantization within the Vilenkin approach.


The reason we concentrate our attention on the Taub
cosmology in the presence of a massless scalar field
consists both of the presence of the necessary Minisuperspace
degrees of freedom and because the WKB construction
of a Schrodinger evolution for the real quantum variables
naturally apply for the Universe anisotropy degree of freedom.
More specifically, on one hand, the possibility to deal
with two quasi-classical variables and a purely quantum one,
allows to fully implement the scheme introduced in \cite{vilenkin}
and, on the other hand, the polymer regularized anisotropy
variable (we will see that its value no longer diverges, as
in standard evolution) is a phase space sample
very appropriate to the concept of "smallness'' of
the quantum system also invoked in \cite{vilenkin}.

Furthermore, the Taub cosmology has a non-trivial
meaning for the physics of the early Universe.
It corresponds to a Bianchi IX model with two scale
factor equal to each other and, it is well-known, that
the Bianchi dynamics in the "corner'' of the spatial
curvature induced potential \cite{montani1} closely resembles small
oscillations around a Taub configuration. Thus
the generality of the Bianchi IX cosmology, versus a generic inhomogeneous 
cosmological model \cite{montani2} justifies the interest for
the present analysis. Finally, implementing the polymer
paradigm within a WKB decomposition of the Minisuperspace
dynamics, we are trying to clarify the behavior
of the anisotropy degree of freedom when a Big-Bounce emerges.
The case of a vacuum Taub cosmology, when the polymer
quantum mechanics is implemented on the anisotropy
dynamics only, was analyzed in \cite{montaniP}, showing
how the cosmological singularity is not removed, but
only probabilistic weakened. The merit of such an investigation
consists in clarifying that the emergence of a bouncing
cosmology requires that the polymer reformulation
also involves the Universe volume. In this respect,
the present analysis is the conceptual continuation
of the study in \cite{montaniP}. We include in the quantum dynamics
a massless scalar field in order to deal with
a relational time variable giving a material nature
\cite{kuchar}.

When we face the description of
the anisotropy degree of freedom as a pure quantum
variable, we adopt the quasi-classical representation
for both the Universe volume and the scalar field.
However, following the analysis in \cite{vilenkin},
we are able to identify, in the end, the label
time coordinate along the space-time slicing,
with a suitable function of the volume and the massless scalar field, by using a gauge
fixing.
All the variable are approached in the polymer formulation
and therefore we are able to infer a bouncing cosmology,
with the very important feature that the anisotropy
degree of freedom is now really "small'' in the
sense of the WKB analysis requirement, see also
\cite{agostini} for a more precise
characterization of this concept.

By means of some non-trivial technicalities,
like a suitable re-definition of the Misner-Chitre'-like
variables here adopted, we finally
demonstrate that the Taub cosmology is
a good candidate, in the present paradigm, for
describing a cyclical anisotropic Universe, always
remaining not to far form the Robertson-Walker geometry.

\section{Conclusion}
We developed a technical algorithm to implement the WKB approach to the quantum Minisuperspace dynamics \cite{vilenkin} within the Polymer representation of quantum mechanics \cite{corichi}. 
One of the difficulties of the analysis above consisted in the necessity to deal with the momenta representation of the quantum dynamics, the only viable for the Polymer quantization procedure, as approached in the continuum limit. The point is that the potential term of the Minisuperspace Hamiltonian is, in general, not quadratic in the Minisuperspace variable, like the kinetic part is in the momenta. 

We proposed a procedure to construct the semi-classical WKB limit in the momentum representation, which is, in principle, applicable to any Minisuperspace system. Such an algorithm has the aim to implement the concept of a cut-off on the quantum dynamics of the Universe, by separating the dynamics into a quasi-classical evolution of a set of configurational variables, e.g. the Universe volume, and those ones rapidly evolving in a fully quantum picture of the dynamics. According to the original idea proposed in \cite{vilenkin}, we arrive to define a Schrodinger-like equation for the quantum subsystem, allowing a consistent interpretation of the wavefunction. 

Then, we applied the general procedure constructed above, to the particular case of a Taub cosmology, as described in the framework of Misner-Chitrè\'e -like variables. We consider as quasi-classical variables the most closely resembling the Universe volume and a suitable function of the free massless scalar field included in the dynamics. As purely quantum variable, we adopt that one most closely resembling the Universe anisotropy. 

As a result, we get a consistent cosmological picture, describing a cyclical Universe in which a quantum anisotropy is regularized, i.e. its amplitude is always finite. The obtained cosmological paradigm is of significant interest in view of constructing a realistic global (quantum and classical) dynamics of the Universe, being characterized by a regular minimum volume turning point (the Big-Bounce), where the possibility for an interpretation of the anisotropy wavefunction can be coherently pursued. Furthermore, such a resulting model has a maximum volume turning point, living in the pure classical region of the dynamics for all configurational coordinates and allowing for the emergence of cyclical closed Universe dynamics, slightly generalizing the positive curved Robertson-Walker geometry, but removing the singular point in which the Big-Bang takes place for the Standard Cosmological Model \cite{montani2, kolb, weinberg}.

\end{document}